\documentclass[conference]{IEEEtran}
\IEEEoverridecommandlockouts
\usepackage{cite}
\usepackage{amsmath,amssymb,amsfonts}
\usepackage{algorithmic}
\usepackage{graphicx}
\usepackage{textcomp}
\usepackage{xcolor}
\usepackage{subfig}

\usepackage{hyperref}
\usepackage{array}
\def\BibTeX{{\rm B\kern-.05em{\sc i\kern-.025em b}\kern-.08em
    T\kern-.1667em\lower.7ex\hbox{E}\kern-.125emX}}
\usepackage{bbold}
\begin{document}

\title{A Sneak Attack on Segmentation of Medical Images Using Deep Neural Network Classifiers\\
}

\author{\IEEEauthorblockN{Shuyue Guan}
\IEEEauthorblockA{\textit{Department of Biomedical Engineering} \\
\textit{The George Washington University}\\
Washington DC, USA \\
frankshuyueguan@gwu.edu \\ \url{https://orcid.org/0000-0002-3779-9368}}
\and
\IEEEauthorblockN{Murray Loew}
\IEEEauthorblockA{\textit{Department of Biomedical Engineering} \\
\textit{The George Washington University}\\
Washington DC, USA \\
loew@gwu.edu}
}
\IEEEoverridecommandlockouts
\IEEEpubid{\makebox[\columnwidth]{\hfill} \hspace{\columnsep}\makebox[\columnwidth]{ }}

\maketitle

\IEEEpubidadjcol

\begin{abstract}
Instead of using current deep-learning segmentation models (like the UNet and variants), we approach the segmentation problem using trained Convolutional Neural Network (CNN) classifiers, which automatically extract important features from images for classification. Those extracted features can be visualized and formed into heatmaps using Gradient-weighted Class Activation Mapping (Grad-CAM). This study tested whether the heatmaps could be used to segment the classified targets. We also proposed an evaluation method for the heatmaps; that is, to re-train the CNN classifier using images filtered by heatmaps and examine its performance. We used the mean-Dice coefficient to evaluate segmentation results. Results from our experiments show that heatmaps can locate and segment partial tumor areas. But use of only the heatmaps from CNN classifiers may not be an optimal approach for segmentation. We have verified that the predictions of CNN classifiers mainly depend on tumor areas, and dark regions in Grad-CAM’s heatmaps also contribute to classification.
\end{abstract}


\section{Introduction}
For image classification, the Convolutional Neural Network (CNN) has performed well in many tasks~\cite{sze_efficient_2017}. Traditional classification methods have relied on manually extracted features; alternatively, the CNN automatically extracts the features from images for classification~\cite{lo_artificial_1995}. To visualize extracted features from a CNN model, recent techniques such as the Grad-CAM~\cite{selvaraju2017grad} can weight and combine the features to display heatmaps of targets on input images, and the targets are the basis of classification. Thus, such techniques provide a way to find the targets of classification from a trained classifier.

Since the CNN has been applied to many medical image classification problems~\cite{bakator_deep_2018}, it will be meaningful if we could gain more knowledge or information about the objects of classification from extracted features of theirs. The development of deep learning, moreover, has made significant contributions to medical image segmentation and become a research focus in the field of medical image segmentation~\cite{su13031224}. Thus, our question is \textit{whether it is possible to segment targets from a trained classifier of the targets.}

\begin{figure}
    \centering
    \begin{tabular}{cc}
 \includegraphics[width=.22\textwidth]{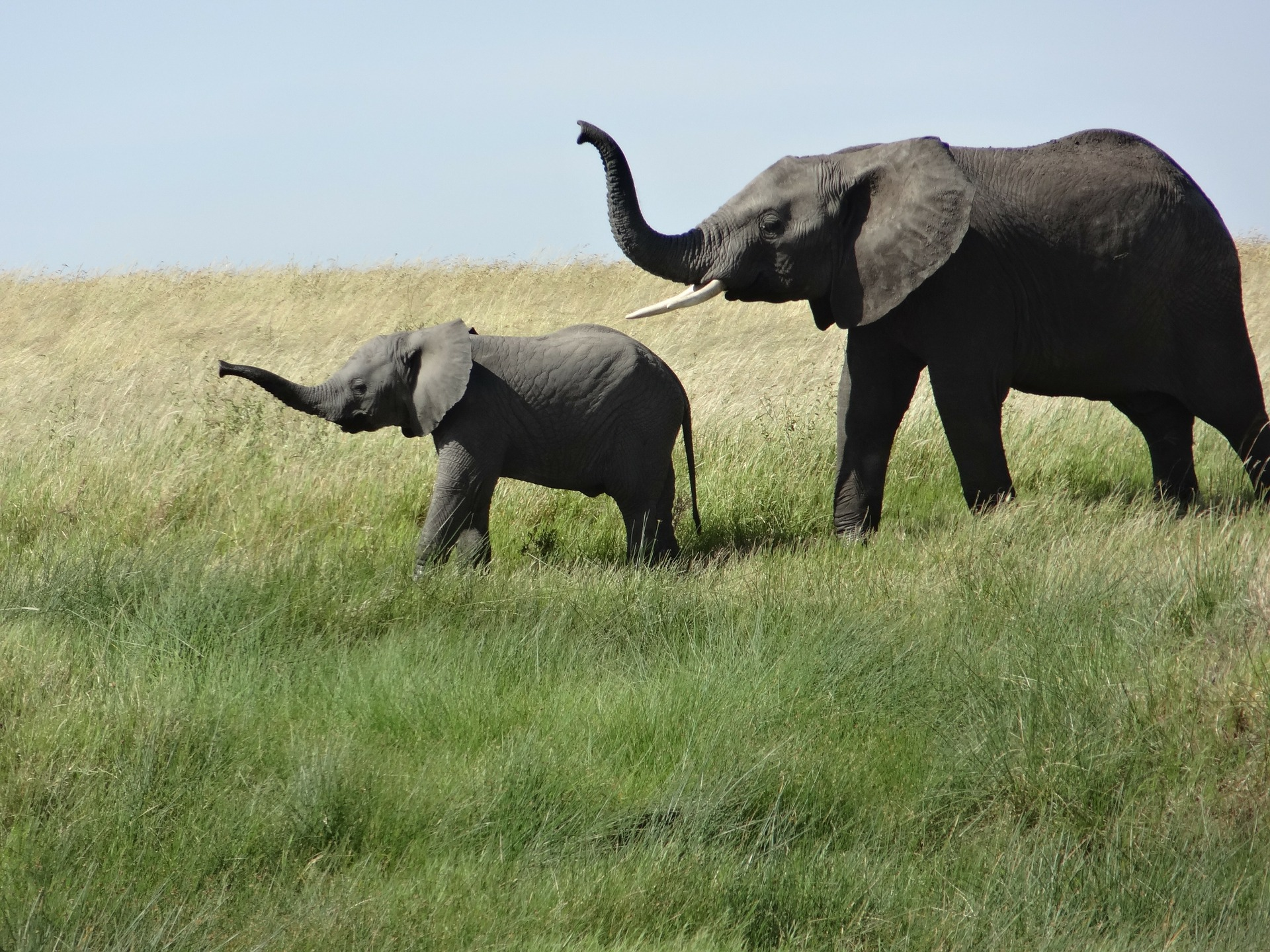} & \includegraphics[width=.22\textwidth]{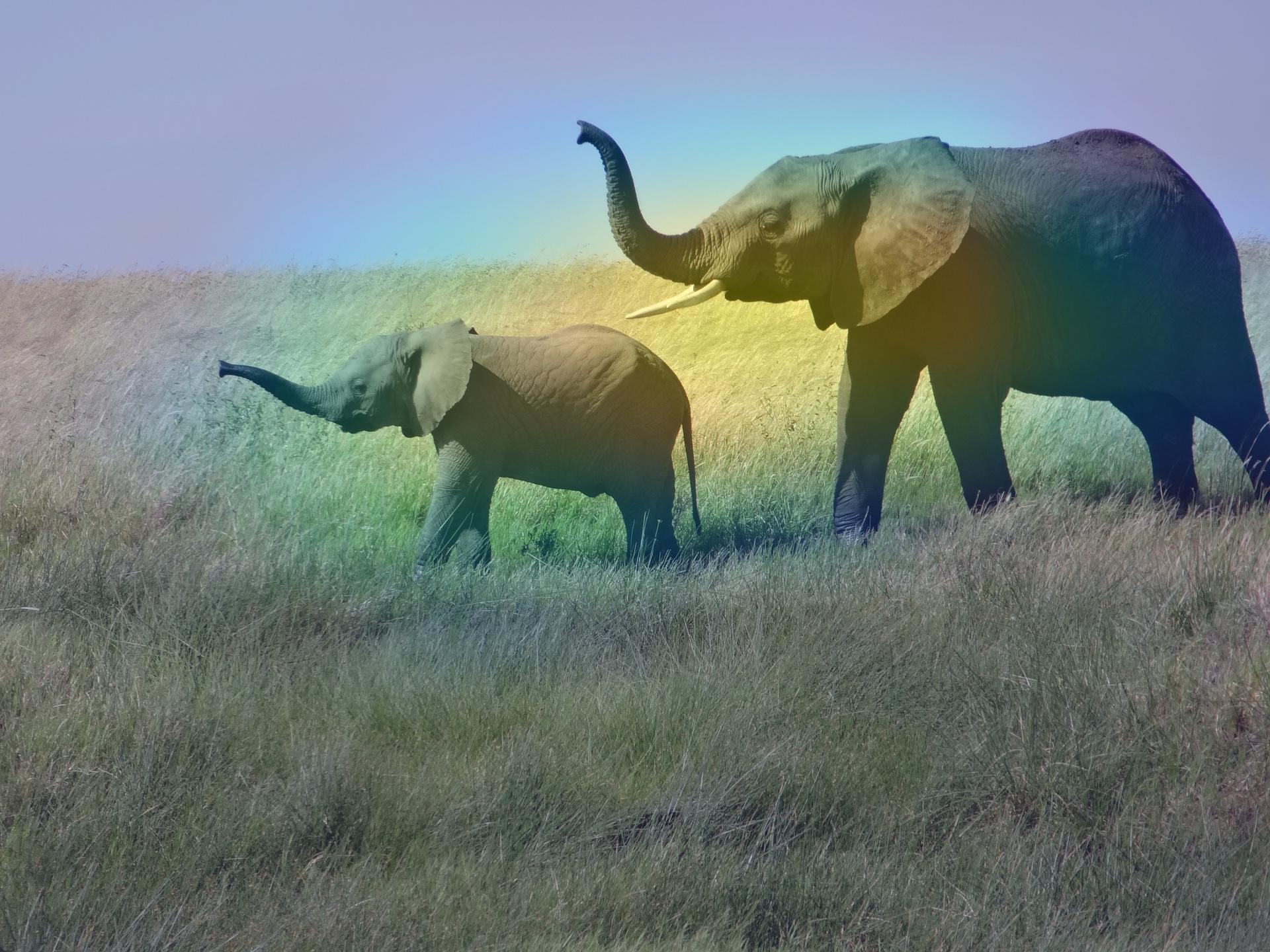}
  \\ 
  (a) & (b)
  \vspace{1em}
  \\
  \includegraphics[width=.22\textwidth]{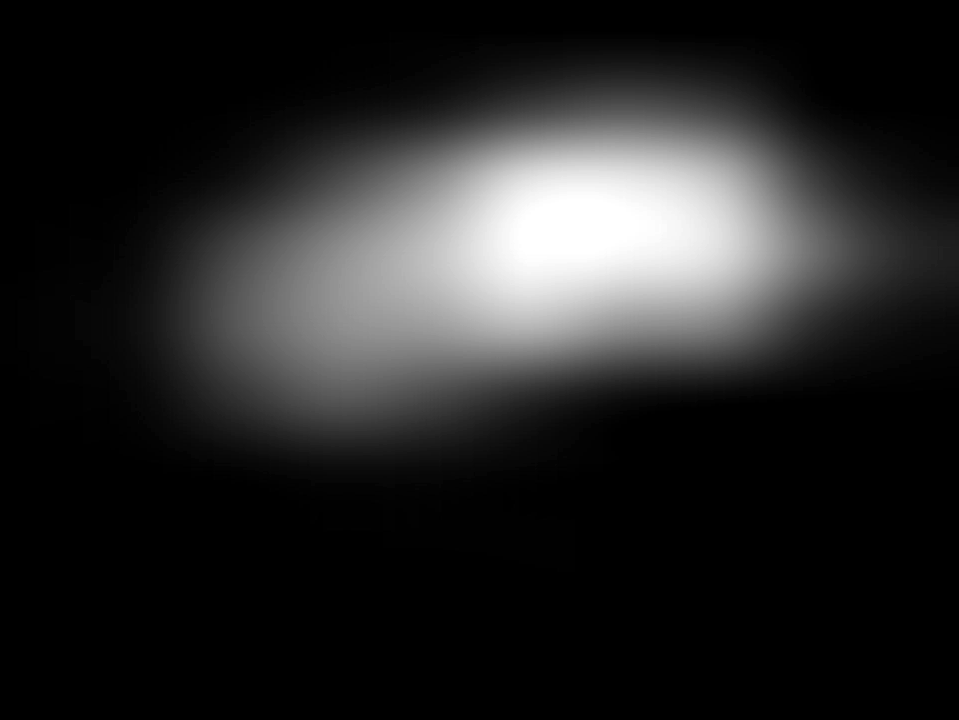} & \includegraphics[width=.22\textwidth]{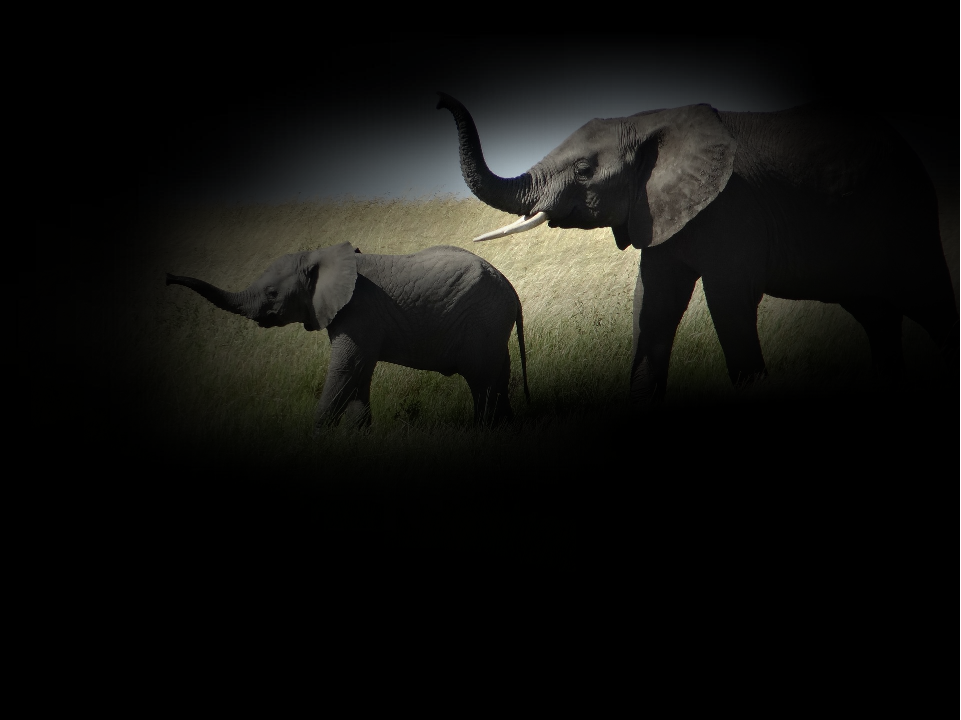} \\
  (c) & (d)
    \end{tabular}
    \caption{Results of Grad-CAM applied to Xception model with input of an elephants image. (a) is the input image. (b) is original image masked by Grad-CAM heatmap (using `\texttt{Parula}' colormap) of the prediction on this input. (c) is the Grad-CAM heatmap mask using gray-scale colormap. (d) is original image filtered by the heatmap mask. \label{fig:exp_elephant}}
\end{figure}

If the features extracted from a trained classifier (CNN) model of targets could be used to segment the targets well, it will benefit medical image segmentation. For example, we could segment breast cancer areas by re-using a classifier for breast cancer detection without training a new segmentation model. A potential advantage of such segmentation is that the objects in classification tasks are often more difficult to segment from the background because their boundaries are not as apparent as in many segmentation tasks.

Figure~\ref{fig:exp_elephant} shows an example of how the heatmap from the Grad-CAM method segments the targets. We input an image of African elephants (Figure~\ref{fig:exp_elephant}a) into the Xception~\cite{chollet_xception_2017} neural network model pre-trained by the ImageNet database~\cite{ILSVRC15}. The pre-trained Xception model can classify images including 1000-class targets\footnote{\url{https://image-net.org/challenges/LSVRC/2014/browse-synsets}}. Its top-1 prediction result of the input image (Figure~\ref{fig:exp_elephant}a) is `\texttt{African\_elephant}'. Then, we apply Grad-CAM to show the heatmap of this prediction; results are shown in Figure~\ref{fig:exp_elephant}. Finally, the input image filtered by the heatmap mask can be considered as a segmentation result of the targets segmented from the background (grass and sky). 

This example shows that we can achieve a segmentation result without training a segmentation model but by using a trained classifier. In this study, we applied this method to mammographic images for breast tumor segmentation. We used breast tumor images from the DDSM database and various CNN-based classifier models (\textit{e.g.}, Xception). Since DDSM describes the location and boundary of each abnormality by a chain-code, we were able to extract the true segmentations of tumor regions. We used the regions of interest instead of entire images to train CNN classifiers. After training the two-class (with- or without-tumor) classifier, we applied Grad-CAM to the classifier for tumor region segmentation. We expect that this will be a beneficial method for general medical image segmentation; \textit{e.g.}, we could segment breast cancer areas by re-using a classifier developed for breast cancer detection without training a new segmentation model from scratch.

\subsection{Related works}
This study -- that to segment breast tumors by re-using trained classifiers -- is inspired by applications of explainability in medical imaging. A main category of explainability methods is attribution-based methods, which are widely used for interpretability of deep learning~\cite{singh_explainable_2020}. The commonly used algorithms of attribution-based methods for medical images are saliency maps~\cite{levy_breast_2016}, activation maps~\cite{van_molle_visualizing_2018}, CAM~\cite{zhou_learning_2016}/Grad-CAM~\cite{selvaraju2017grad}, Gradient~\cite{eitel_testing_2019}, SHAP~\cite{young_deep_2019}, \textit{et cetera}.

In this study, we examined how the attribution maps (the heatmaps) generated from the Grad-CAM algorithm can contribute to the segmentation of breast tumors. The visualization of the class-specific units~\cite{zhou_object_2015, zhou_learning_2016} for CNN classifiers is used to locate the most discriminative components for classification in the image. The authors of Grad-CAM also evaluated the localization capability of Grad-CAM~\cite{selvaraju2017grad} by bounding boxes containing the objects. But those methods provide coarse boundaries around the targets, and these studies have not provided further quantitative analysis about the differences between predicted and real boundaries of the targets. Thus, they are considered to be methods for localization rather than segmentation. For weakly-supervised image segmentation~\cite{kolesnikov_seed_2016,selvaraju2017grad}, CAM/Grad-CAM's heatmaps can be computed and combined with other segmentation models, such as UNet CNNs~\cite{nguyen_novel_2019, rajapaksa_localized_2021}, to improve segmentation performance. We are aware of no similar study that has merely applied the Grad-CAM algorithm by trained CNN classifiers to a specific application of medical image segmentation, without using any other segmentation models or approaches. We quantitatively analyzed the differences between predicted boundaries from Grad-CAM and real boundaries of the targets, and we discussed the relationships between the performance of segmentation and classification based on the CNN classifier.

\section{Methods}

\begin{figure*}[t]
    \centering
    \includegraphics[width=0.9\textwidth]{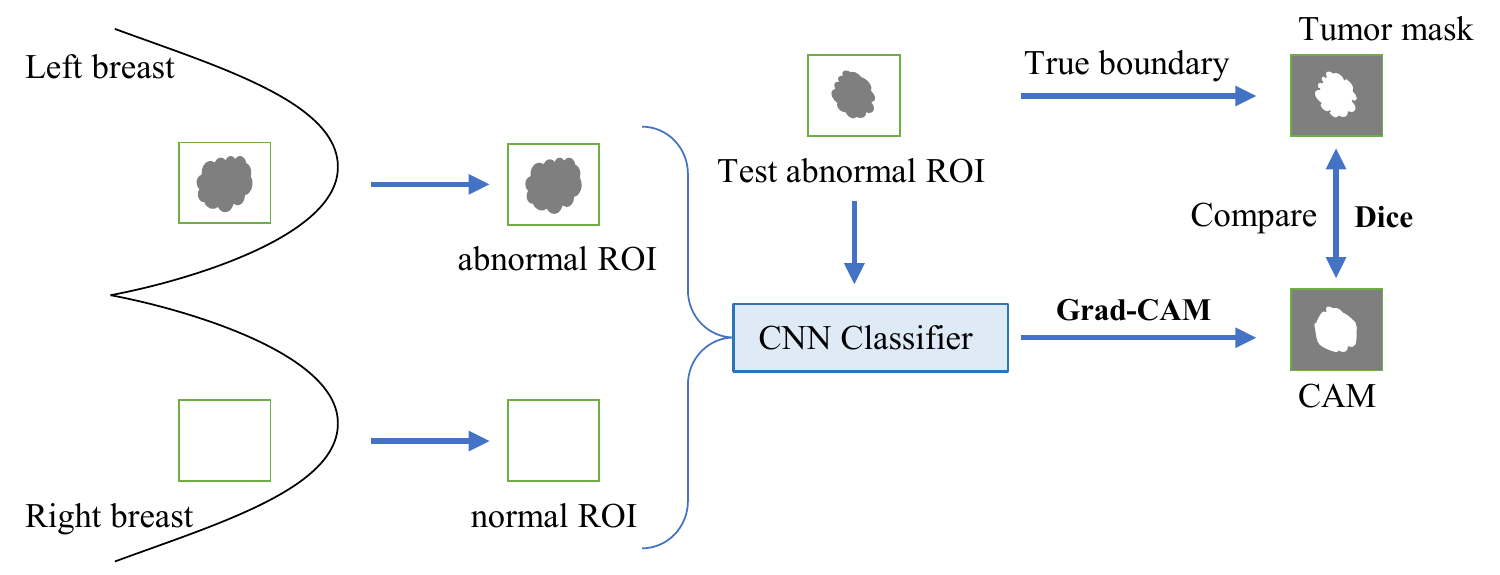}
    \caption{Flowchart of the Experiment \#1. The true boundaries of tumor regions in abnormal ROIs are provided by the DDSM database. \label{fig:exp-plan-1}}
\end{figure*}

\subsection{Grad-CAM method}
To compute the class activation maps (CAM), Zhou \textit{et al.}~\cite{zhou_learning_2016} proposed to insert a global average pooling (GAP) layer between the last convolutional layer (feature maps) and the output layer in CNNs. The size of each 2-D feature map is $[x,y]$, and a value in the $k$-th feature map is $f_{i,j}^k$.
Suppose the last convolutional layer contains $n$ feature maps, then the GAP layer will have $n$ nodes for each extracted feature map. From the definition, the value of the $k$-th node is the average value of the $k$-th feature map:
\begin{equation} \label{eq:fk}
    F^k=\frac{1}{xy}\sum_{\substack{i\in \{1,2,\cdots, x\} \\ j\in \{1,2,\cdots, y\}}} f_{i,j}^k
\end{equation}

After inserting the GAP layer, to obtain the weights of connections from the GAP layer to the output layer requires to re-train the whole network with training data. $w_k^c$ is the weight of connection from the $k$-th node in the GAP layer to the $c$-th node in the output layer (for the $c$-th class). Thus, the CAM of $c$-th class is:
\begin{equation} \label{eq:cam}
\text{CAM}^c[i,j]=\sum_{k=1}^{n} w_k^c \cdot f_{i,j}^k
\end{equation}
The size of CAM is the same as that of the feature maps.

The authors (Selvaraju \textit{et al.}~\cite{selvaraju2017grad}) of Grad-CAM noted that re-training is not necessary. They applied the back-propagation gradient and chain rule to calculate the connection weights $w_k^c$ instead of re-training.

The value of the $c$-th node in output layer $Y^c$ is the score of the target classification for the $c$-th class:
\begin{equation}
    Y^c=\sum_{k=1}^{n} w_k^c \cdot F^k
\end{equation}
By taking the partial derivative of $F^k$:
\begin{equation}
    \frac{\partial Y^c}{\partial F^k}=w_k^c
\end{equation}
By the chain rule:
\begin{equation} \label{eq:chain-rule}
    w_k^c=\frac{\partial Y^c}{\partial F^k}=\frac{\partial Y^c}{\partial f_{i,j}^k}\cdot \frac{\partial f_{i,j}^k}{\partial F^k}
\end{equation}
From the Equation~\ref{eq:fk}, by taking the partial derivative of $f_{i,j}^k$:
\begin{equation} \label{eq:partial-fk}
    \frac{\partial F^k}{\partial f_{i,j}^k}=\frac{1}{xy}
\end{equation}
Therefore, by taking Equations~\ref{eq:chain-rule}~and~\ref{eq:partial-fk} to eliminate the $F^k$:
\begin{equation} \label{eq:weights}
    w_k^c=\frac{\partial Y^c}{\partial f_{i,j}^k}\cdot xy
\end{equation}
Finally, putting Equations~\ref{eq:weights}~and~\ref{eq:cam} together, they find the way to compute the CAM for $c$-th class without really inserting and training a GAP layer. Thus, it is called Grad-CAM:
\begin{equation} \label{eq:grad-cam}
\text{Grad-CAM}^c[i,j]=xy\cdot \sum_{k=1}^{n} \frac{\partial Y^c}{\partial f_{i,j}^k}\cdot f_{i,j}^k
\end{equation}

The size of CAM (Grad-CAM) equals the size of feature maps: $[x,y]$, which is usually smaller than the size of input images. For comparison, resizing is commonly applied to CAMs to enlarge their size to be the same as input images.

\subsection{Proposed experiments} \label{sec:exp-plan}

We have two experiments. In the first experiment, we used the regions of interest (ROIs) of breast cancer images from the DDSM~\cite{heath_digital_2000} to train two-class (with/without tumors) CNN classifiers. ROIs with tumors are called abnormal ROIs and ROIs without tumors are called normal ROIs. After training the two-class classifier using these normal and abnormal ROIs, we will apply Grad-CAM and the classifier to test abnormal ROIs to segment tumor regions. Then, we will use the true boundary of the test whether abnormal ROIs can be used to evaluate segmentation results. Figure~\ref{fig:exp-plan-1} shows the flowchart of this experiment. The goal of the Experiment \#1 is to verify how well the medical targets are segmented by a trained classifier using Grad-CAM algorithm.

\begin{figure}[h]
    \centering
    \includegraphics[width=0.45\textwidth]{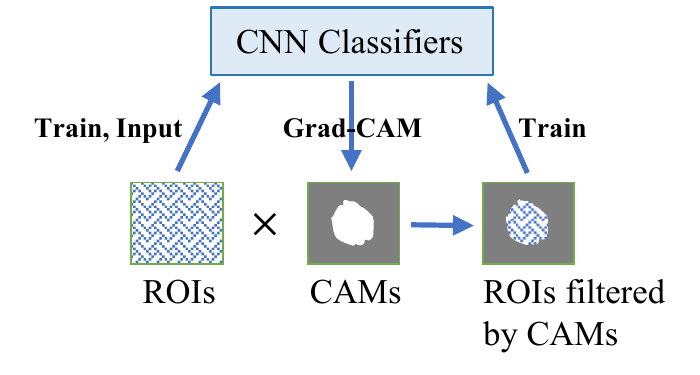}
    \caption{Flowchart of the Experiment \#2. The normal and abnormal ROIs are used twice to train the CNN classifiers and then to generate CAMs by trained classifiers using Grad-CAM algorithm. The CNN classifiers to be trained by CAM-filtered ROIs are the same CNN models (same structures) as trained by the original ROIs before but trained from scratch again.  \label{fig:exp-plan-2}}
\end{figure}

By using the Grad-CAM algorithm, the trained CNN classifiers can generate CAMs from both normal and abnormal ROIs. These CAMs can be considered as masks that indicate the areas that are important to classification. In the second experiment, we trained CNN classifiers from scratch by only using information of those areas. The training data are ROIs filtered by CAMs. It is an evaluation method for the CAMs: to re-train the CNN classifiers using images filtered by CAMs and examine their performance. By combining with the Experiment \#1, the steps of Experiment \#2 are (Figure~\ref{fig:exp-plan-2}):
\begin{itemize}
    \item To train two-class classifiers with normal and abnormal ROIs.
    \item To generate CAMs by inputting these ROIs in trained classifiers using Grad-CAM algorithm.
    \item To create CAM-filtered ROIs: resize CAMs (heatmaps) to the same size of ROIs and convert their range of values to $[0,1]$; then, multiplied by the original ROIs. The important areas to classification are close to 1 in CAMs thus, they will be kept in CAM-filtered ROIs.
    \item To train the same two-class classifiers (same models) from scratch again by CAM-filtered ROIs.
\end{itemize}
The goals of the Experiment \#2 are to examine 1) whether Grad-CAM can really recognize the areas in the images that are important to classification; 2) whether the predictions of CNN classifiers really depend on tumor areas. 

\textbf{For comparison, we additionally trained these CNN classifiers by the truth-mask-filtered ROIs}. To create truth-mask-filtered ROIs is similar to generating the CAM-filtered ROIs. For abnormal ROIs, we multiply ROIs by their corresponding tumor masks so that only the tumor areas are kept and the background (non-tumor area) is removed (pixel values $=0$). For normal ROIs, since there is no tumor area, we multiply ROIs by randomly selected tumor masks from abnormal cases, for the purpose of making normal/abnormal ROIs have similar shapes (outlines).

\section{Materials}

\subsection{Mammography dataset}
Mammography is the process of using low-energy X-rays to examine the human breast for diagnosis and screening. There are two main angles to get the X-ray images: the cranio-caudal (CC) view and the mediolateral-oblique (MLO) view. In this study, we used mammographic images from the Digital Database for Screening Mammography (DDSM)~\cite{heath_digital_2000}. 
The DDSM database contains approximately 2620 cases in 16-bit gray-value: 695 normal cases, 1925 abnormal cases (914 malignant/cancers cases, 870 benign cases and 141 benign without callback) with locations and boundaries of abnormalities. Each case includes four images representing the left and right breasts in CC and MLO views. 

\subsection{Region of Interest (ROI)}

We firstly downloaded mammographic images from DDSM database and cropped the Region of Interest (ROI) by given abnormal areas as ground truth information. Images in DDSM are compressed in LJPEG format. To decompress and convert these images, we used the DDSM Utility~\cite{Anmol2015}. We converted all images in DDSM to PNG format. DDSM describes the location and boundary of actual abnormality by chain-codes, which are recorded in OVERLAY files for each breast image containing abnormalities. The DDSM Utility also provides the tool to read boundary information and display them for each image having abnormalities. Since the DDSM Utility tools run on MATLAB, we implemented all pre-processing tasks using MATLAB.

\begin{figure}[h]
    \centering
    \includegraphics[width=0.4\textwidth]{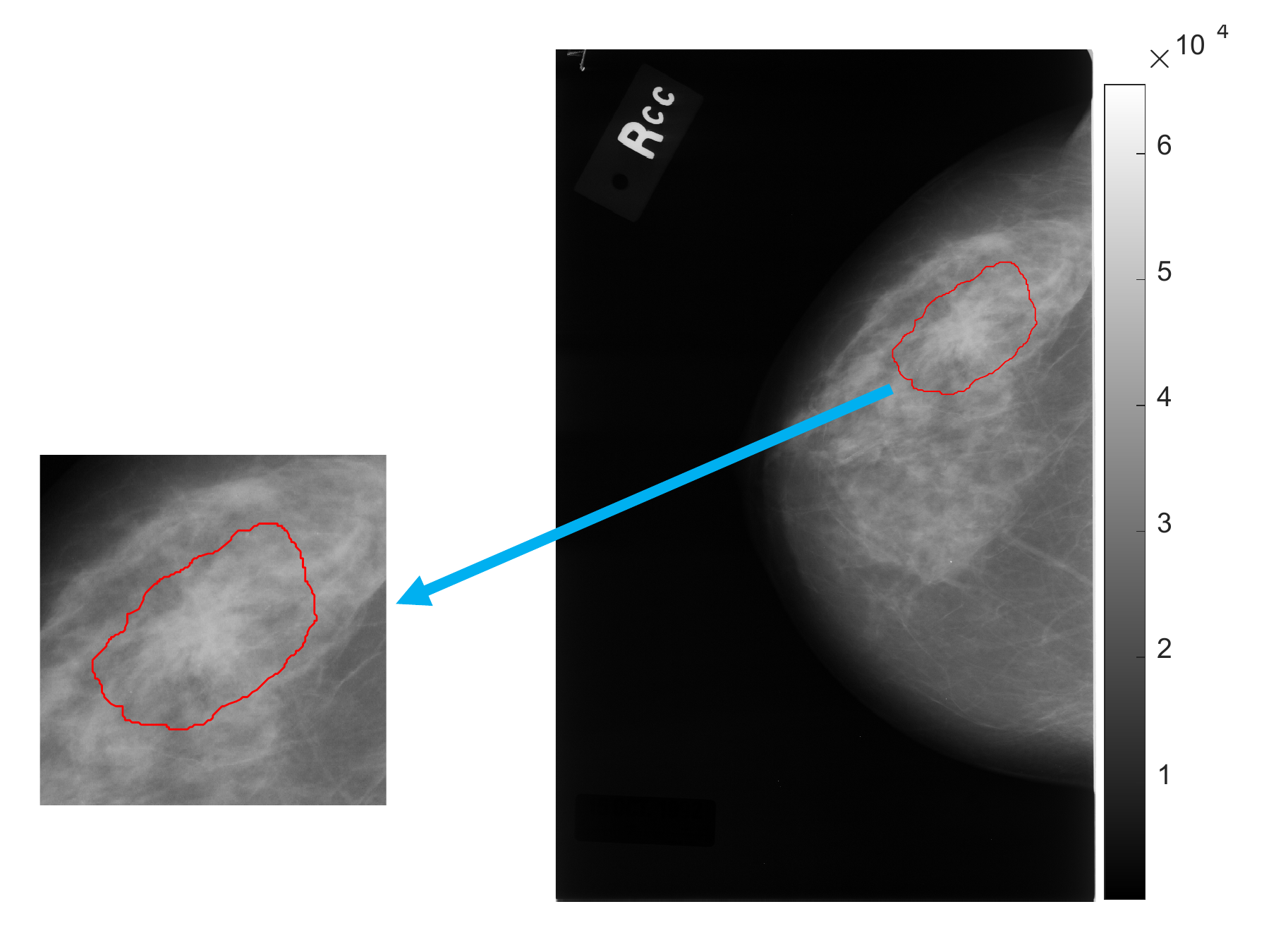}
    \caption{The ROI (left) is cropped from an original image (right) from DDSM dataset. The red boundary shows the tumor area. The ROI is larger than the size of tumor area because of padding. \label{fig:crop-roi}}
\end{figure}

We used the ROIs instead of entire images to train CNN classifiers. These ROIs are cropped rectangle-shape images (Figure~\ref{fig:crop-roi}) and obtained by:
\begin{itemize}
    \item For abnormal ROIs from images containing abnormalities, they are the minimum rectangle-shape areas surrounding the whole given ground truth boundaries \textbf{with padding}.
    
    \item For normal ROIs, they were cropped on the other side of a breast having abnormal ROI and the normal ROI was the same size (\textbf{with padding}) and location as the abnormal ROI on different breast side. If both left and right breasts having abnormal ROIs and their locations overlapping, we discarded this sample. Since in most cases, only one side of breast has tumor and the area and shape of left and right breast are similar; thus, normal ROIs and abnormal ROIs have similar black background areas and scaling. 
    
    \item All ROIs are converted to 8-bit gray-value.
    \item All ROIs are only from the CC views.
\end{itemize}

\begin{figure}[h]
    \centering
    \includegraphics[width=0.25\textwidth]{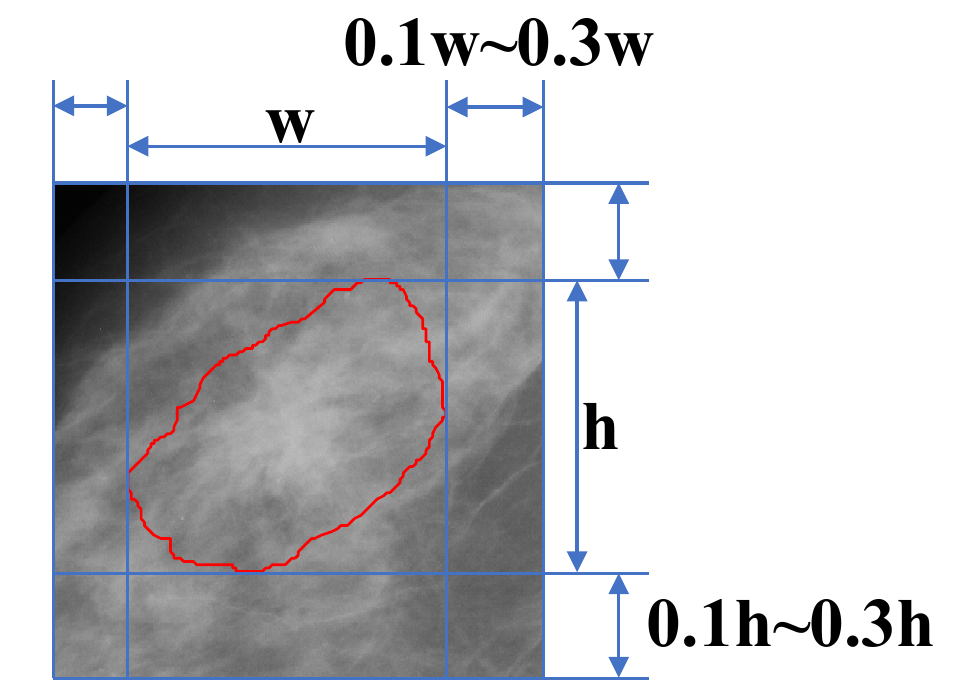}
    \caption{The padding is added to four sides of ROIs by some randomness and depended on the size of tumor area.  \label{fig:padding}}
\end{figure}

The padding is added to all ROIs in order to vary the locations of tumors in abnormal ROIs and to avoid excessive proportion of the tumor area in a ROI. ROIs are larger than the sizes of tumor areas because of padding. As shown in Figure~\ref{fig:padding}, the padding is added by some randomness and depended on the size of tumors:
\begin{itemize}
    \item Width: randomly adding 10\%-30\% of tumor width on left and right sides.
    \item Height: similar as width.
\end{itemize}

\begin{figure}[h]
    \centering
    \includegraphics[width=0.45\textwidth]{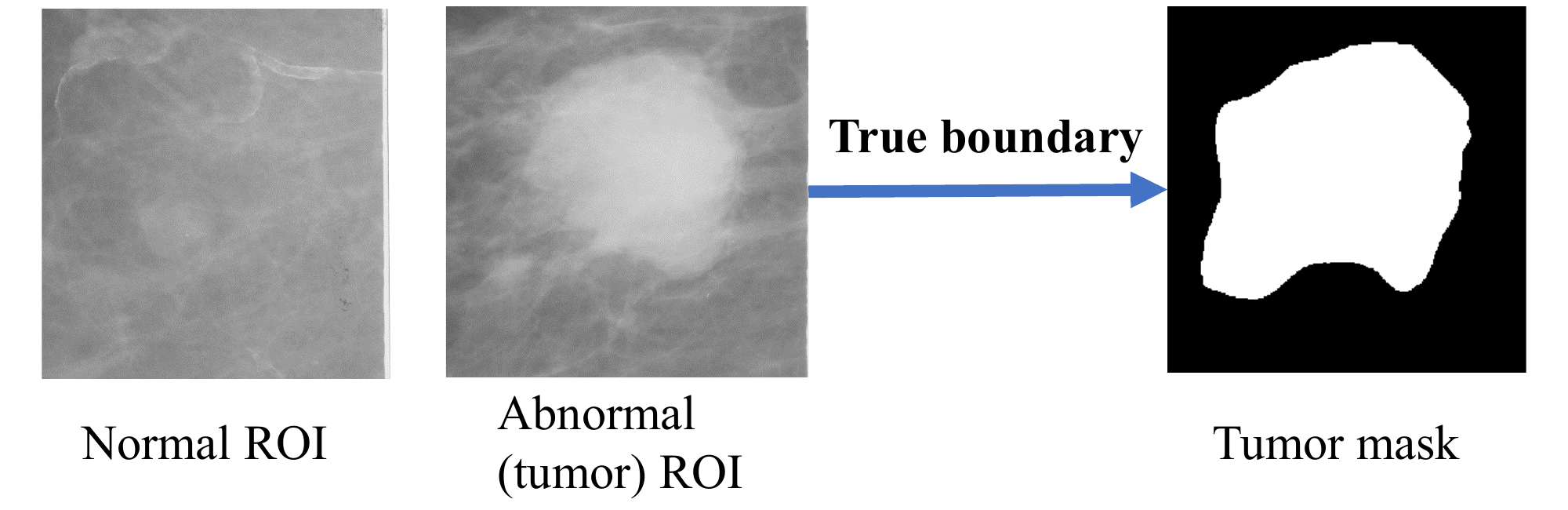}
    \caption{Examples of ROIs. The tumor mask is binary image created from the tumor ROI and truth boundary of the tumor area.  \label{fig:rois}}
\end{figure}

After collecting ROIs, as shown in Figure~\ref{fig:rois}, we have normal ROIs and abnormal (tumor) ROIs to apply classification (using binary labels), and could have real tumor masks to apply segmentation.

\subsection{Classifier models}
To train CNN classifiers by normal and abnormal ROIs and to generate CAMs using Grad-CAM algorithm, we used six CNN models. They are: NASNetMobile~\cite{zoph_learning_2018}, MobileNetV2~\cite{sandler_mobilenetv2_2019}, DenseNet121~\cite{huang_densely_2018}, ResNet50V2~\cite{he_identity_2016}, Xception~\cite{chollet_xception_2017}, and InceptionV3~\cite{szegedy_rethinking_2015}.

\section{Experiments and results}
Except for cropping ROIs, experiments are implemented by codes written in the Python language\footnote{Papers, codes and additional material can be found at the first author’s website linked up with the ORCID: \url{https://orcid.org/0000-0002-3779-9368}}.

\subsection{Training CNN classifiers}
Our dataset has 325 abnormal (tumor) ROIs and 297 normal ROIs in total. To train the CNN classifiers, we divide the dataset into 80\% for training and 20\% for validation. The framework of deep learning models is Keras\footnote{\url{https://keras.io/api/applications/}}. Every CNN model is trained about 200 epochs with \texttt{EarlyStopping}\footnote{\url{https://keras.io/api/callbacks/early\_stopping/}} setting. The classifier models having the best validation (accuracy) performance during each training were saved.

\subsection{Obtaining CAMs and comparison}
To input an abnormal ROI into the trained CNN classifier by using Grad-CAM algorithm, we can obtain a CAM for that ROI. Then, we resized the CAM to the size same as the input ROI. The CAMs are gray-value image, and the truth tumor masks we have are binary image. Thus, we applied the mean-Dice metric to compare CAMs and tumor masks.

The Dice coefficient~\cite{dice1945measures,sorensen1948method} of two binary images $A$ and $B$ is:
\[
\text{Dice}(A,B) = 2\times \frac{|A\cap B|}{|A|+|B|}
\]
To calculate Dice coefficient requires both images are binary; thus, we need to transform the CAMs from gray-value ($[0,255]$) to binary ($\{0,1\}$). Suppose $B$ is the CAM, it can be binarized by setting a threshold ($t$): $B_t(B>t)=1$ and $B_t(B\leq t)=0$, where $B_t$ is the binarized CAM. Then, the mean-Dice metric is defined:
\begin{equation} \label{eq:mean-dice}
    \text{mean-Dice}(A,B) = \frac{1}{256} \cdot \sum_{t=0}^{255} \text{Dice}(A,B_t)
\end{equation}

\subsection{Result of Experiment \#1}
\begin{figure}[h]
    \centering
    \includegraphics[width=0.4\textwidth]{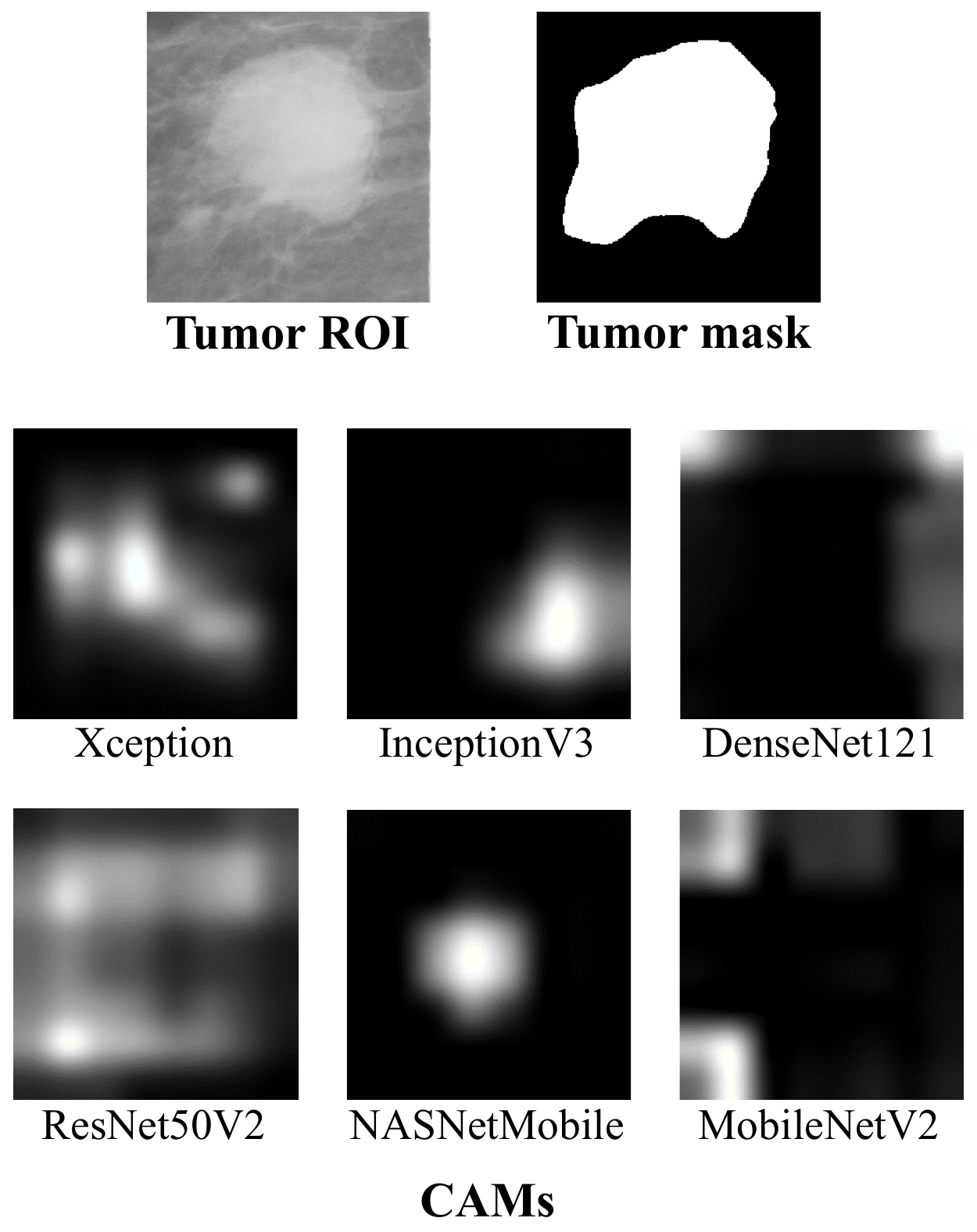}
    \caption{Result of Experiment \#1. The first row shows one of the abnormal (tumor) ROIs and its truth mask. Other rows show the CAMs of this ROI generated by using trained CNN classifiers and Grad-CAM algorithm. \label{fig:res1}}
\end{figure}

We report the best validation accuracy (val\_acc) and averaged mean-Dice (Dice) of Experiment \#1 (described in Section~\ref{sec:exp-plan} and Figure~\ref{fig:exp-plan-1}) for each CNN classifier in Table~\ref{tab:exp1-res}. The averaged mean-Dice is calculated using all 325 abnormal (tumor) ROIs. Figure~\ref{fig:res1} shows CAMs of one tumor ROI generated by using trained CNN classifiers and Grad-CAM algorithm. 

As shown in the result, CAMs from Xception overlap the most regions of true tumor masks. But CAMs from DenseNet121 and MobileNetV2 almost do not cover the true tumor regions. We could see from Figure~\ref{fig:res1}, heatmaps (CAMs) of the two classifiers highlight the corners and outer areas of images instead of the tumor regions. Although CAMs from DenseNet121 and MobileNetV2 have very small Dice values with true tumor areas, they still have good classification performance. Thus, the result leads to two questions:
\begin{enumerate}
    \item Can Grad-CAM really recognize the important areas in the images to classification?
    \item Do the predictions of CNN classifiers really depend on tumor areas?
\end{enumerate}
Experiment \#2 is proposed to examine the two questions.

\begin{table}[htbp]
\caption{Result of Experiment \#1. Descending sort by val\_acc.}
\begin{center}
\begin{tabular}{|p{0.12\textwidth}
|>{\centering\arraybackslash}p{0.08\textwidth}|>{\centering\arraybackslash}p{0.08\textwidth}|}
\hline
\textbf{Classifier} & \textbf{val\_acc} & \textbf{Dice}\\ \hline
InceptionV3 & 0.872 & 0.256 \\ \hline
DenseNet121$^{\mathrm{a}}$ & 0.872 & 0.030 \\ \hline
Xception & 0.856 & \textbf{0.435} \\ \hline
NASNetMobile & 0.848 & 0.353 \\ \hline
MobileNetV2$^{\mathrm{a}}$ & 0.840 & 0.034 \\ \hline
ResNet50V2 & 0.840 & 0.365 \\ \hline
\multicolumn{3}{l}{$^{\mathrm{a}}$These classifiers have very small Dice values.}
\end{tabular}
\label{tab:exp1-res}
\end{center}
\end{table}

\subsection{Result of Experiment \#2}
\begin{figure}[h]
    \centering
    \includegraphics[width=0.45\textwidth]{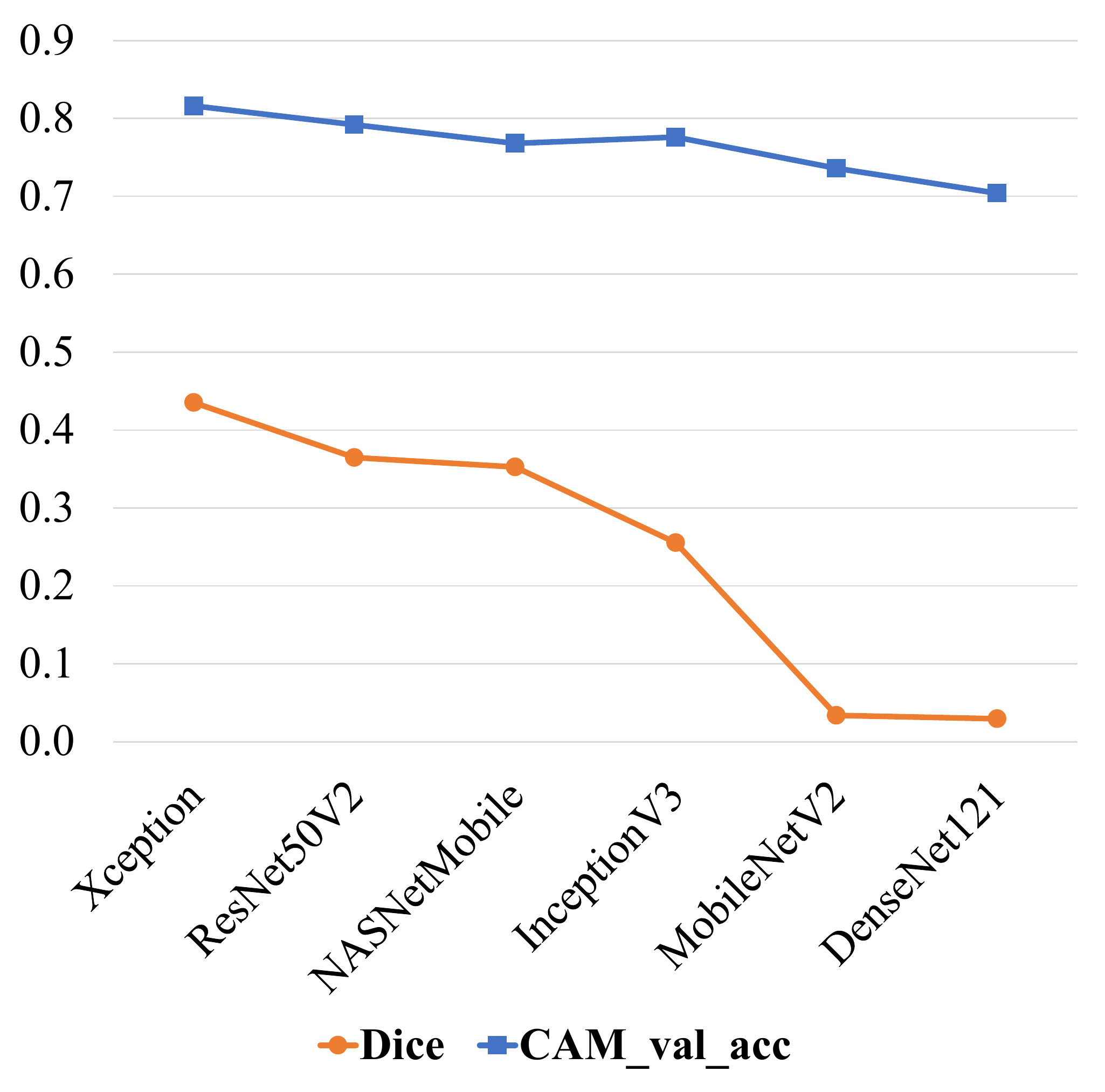}
    \caption{Plots of Dice and CAM\_val\_acc for the six CNN classifiers in Table~\ref{tab:exp2-res}.  \label{fig:res2}}
\end{figure}

In addition, we add the best validation accuracy of CNN classifiers trained by the CAM-filtered ROIs (CAM\_val\_acc) and by truth-mask-filtered ROIs (mask\_val\_acc), which are described in Section~\ref{sec:exp-plan} and Figure~\ref{fig:exp-plan-2}, to the result in Table~\ref{tab:exp2-res}. It is extended from Table~\ref{tab:exp1-res} and descending sort by Dice. And, we plot the Dice and CAM\_val\_acc for the six CNN classifiers in Figure~\ref{fig:res2}.

\begin{table}[htbp]
\caption{Result of Experiment \#2. Descending sort by Dice.}
\begin{center}
\begin{tabular}{|l|c|c|c|c|}
\hline

\textbf{Classifier} & \textbf{val\_acc} & \textbf{Dice} & \textbf{CAM\_val\_acc} & \textbf{mask\_val\_acc} \\ \hline
Xception & 0.856 & 0.435 & \textbf{0.816} & 0.880 \\ \hline
ResNet50V2 & 0.840 & 0.365 & 0.792 & 0.880 \\ \hline
NASNetMobile & 0.848 & 0.353 & 0.768 & 0.872 \\ \hline
InceptionV3 & \textbf{0.872} & 0.256 & 0.776 & \textbf{0.904} \\ \hline
MobileNetV2 & 0.840 & 0.034 & 0.736 & 0.872 \\ \hline
DenseNet121 & \textbf{0.872} & 0.030 & 0.704 & 0.896 \\ \hline

\end{tabular}
\label{tab:exp2-res}
\end{center}
\end{table}

As shown in Figure~\ref{fig:res2}, in general, training on ROIs filtered by CAMs covering more tumor areas (higher Dice values) leads to better classification performance (CAM\_val\_acc). InceptionV3 model is an exception: its Dice is smaller than NASNetMobile's but it has a higher CAM\_val\_acc than NASNetMobile. The reason may be that InceptionV3 has a better classification capability than NASNetMobile because 1) the parameters in InceptionV3 are about five times of the parameters in NASNetMobile\footnote{\url{https://keras.io/api/applications/}}; 2) Table~\ref{tab:exp2-res} shows InceptionV3 has higher val\_acc and mask\_val\_acc than NASNetMobile.

\begin{figure}[h]
    \centering
    \begin{tabular}{cc}
 \includegraphics[width=.12\textwidth]{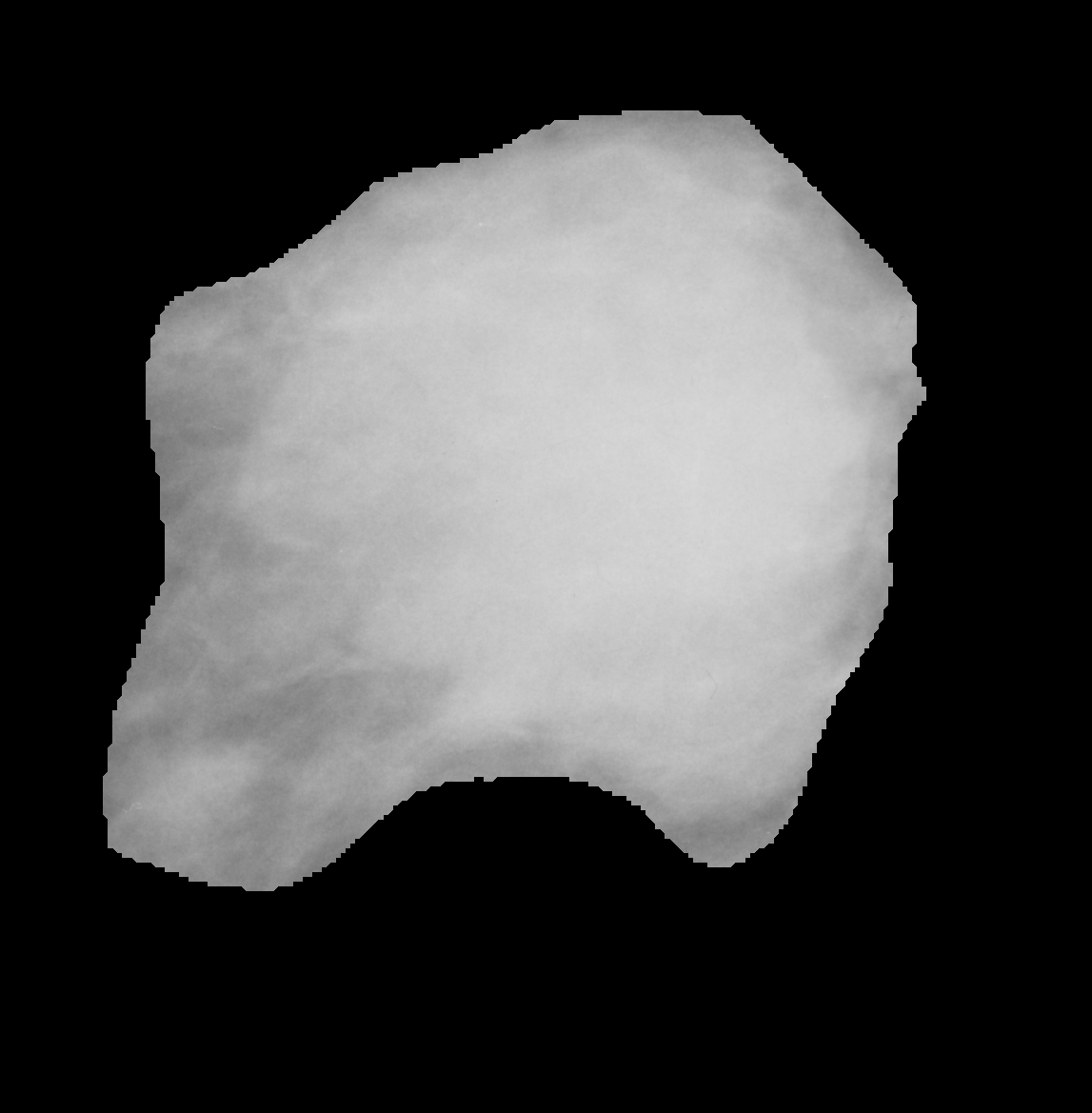} & \includegraphics[width=.12\textwidth]{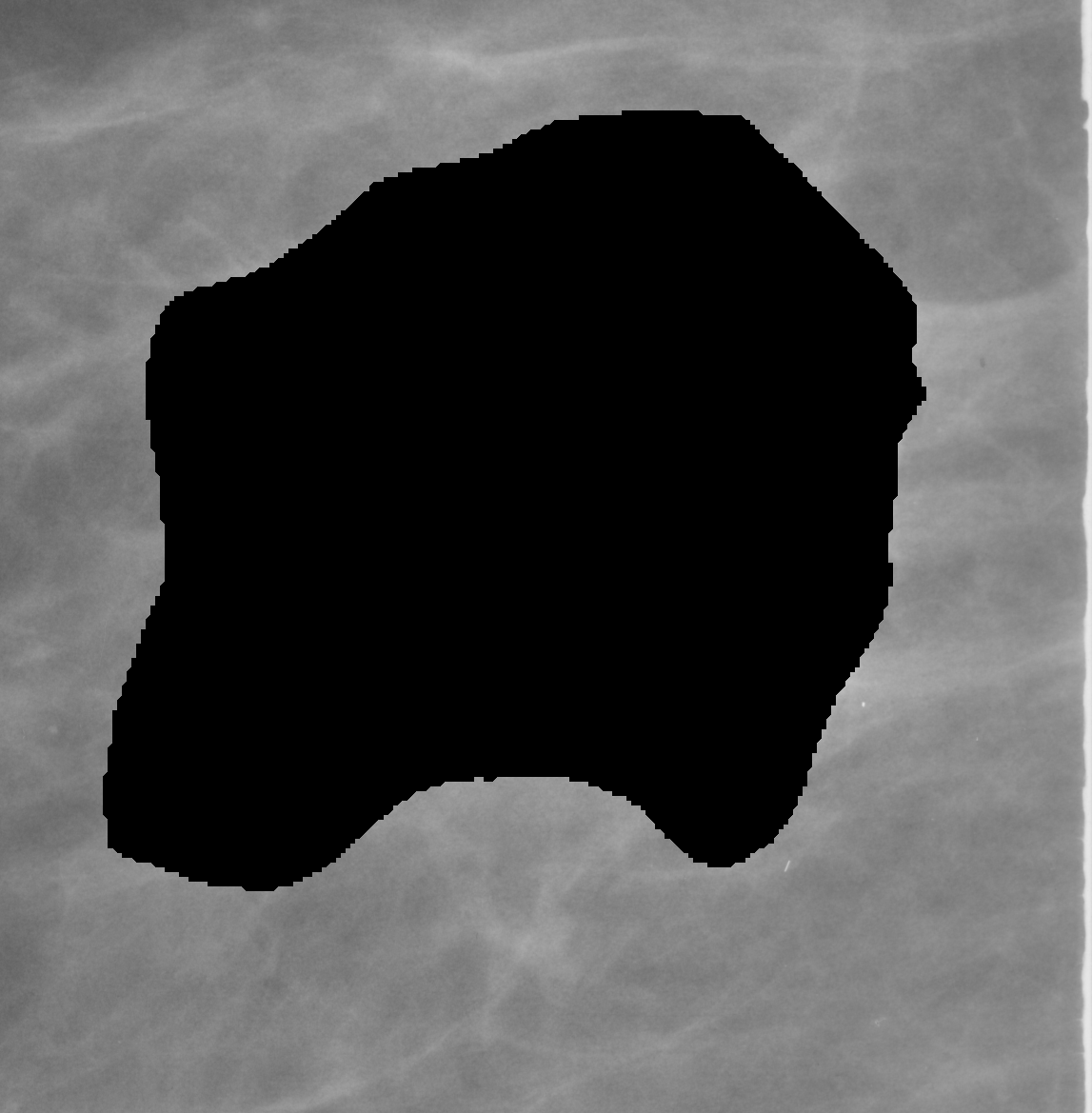}
  \\ 
  (a) & (b)
  \vspace{1em}
    \end{tabular}
    \caption{Examples of truth-mask-filtered (a) and  inverse-mask-filtered (b) ROIs from the case shown in Figure~\ref{fig:res1}.\label{fig:mask-filtered}}
\end{figure}

For all CNN classifiers, training on ROIs filtered by truth tumor masks (Figure~\ref{fig:mask-filtered}a) leads to the best classification performance (Table~\ref{tab:exp2-res}, mask\_val\_acc), even better than training on original ROIs. Using truth-mask-filtered ROIs is better than CAM-filtered ROIs simply because truth-mask-filtered ROIs contain the whole tumor areas (Dice $=1$). And, the reason that using truth-mask-filtered ROIs is better than original ROIs might be that original ROIs contain some irrelative information interrupting classification. Such result indicates that \textit{the classification depends on tumor areas more than other areas}. To confirm this conclusion, we trained InceptionV3 on ROIs filtered by \textbf{inverse} truth tumor masks (Figure~\ref{fig:mask-filtered}b). The inverse-mask-filtered ROIs exactly exclude the tumor areas (Dice $=0$). The best validation accuracy of InceptionV3 trained by inverse-mask-filtered ROIs is \textbf{0.616}. Comparing with the CAM\_val\_acc (0.776), val\_acc (0.872), and mask\_val\_acc (0.904) of InceptionV3, ROIs containing smaller tumor areas lead to worse classification performance.

\section{Discussions}
\begin{figure}[h]
    \centering
    \includegraphics[width=0.48\textwidth]{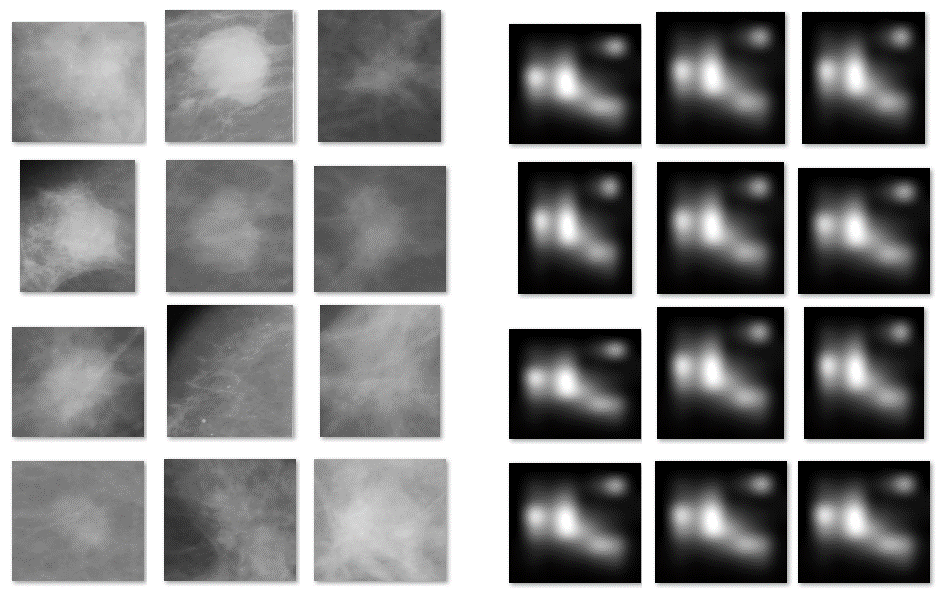}
    \caption{Some tumor ROIs and their CAMs from Xception.  \label{fig:xception-cams}}
\end{figure}

The results of Experiment \#1 show that the best segmentation from Grad-CAM method is 0.435 (averaged mean-Dice) by the Xception model. Such results indicate CAMs may not perform segmentation well for binary-classification problems because the highlighted area (hot region) on the CAM is almost the same for all CAMs (Figure~\ref{fig:xception-cams}). The CNN classifiers may select some fixed areas that are best for classification. Thus, these classifiers achieve a good classification performance but their CAMs are not optimal for segmentation. In addition, the requirements of segmentation and classification are different: segmentation is based on local (pixel-wise) decisions but classification is based on a global decision. The decision of classification may depend on part of the target-object and/or other parts outside of the target-object. 
Therefore, only use of CAMs from CNN classifiers may not be an optimal approach for segmentation; instead, to combine CAMs with other segmentation methods can be a promising direction~\cite{nguyen_novel_2019, rajapaksa_localized_2021}.

Experiment \#2 verified that the predictions of CNN classifiers mainly depend on tumor areas. The Grad-CAM algorithm can recognize some of the important (tumor) areas to classification but its performance depends on classifier models. As shown by the DenseNet121 in Table~\ref{tab:exp2-res} and Figure~\ref{fig:res1}, its CAMs have very small Dice values with true tumor areas but the model still has a good classification performance. This implies that the dark regions in CAMs also contribute to classification.

\subsection{Future works}
This study may raise other questions and discover the starting points for future studies to make progress in the understanding of deep learning. The Grad-CAM is not the only method to generate the heatmap that reflects the basis of classification. In future works, we would test other techniques, such as Saliency map~\cite{levy_breast_2016}, SHAP~\cite{young_deep_2019}, and Activation map~\cite{van_molle_visualizing_2018}, to create segmentation and make comparison. We found the dark-regions in CAMs from Grad-CAM also contribute to classification; thus, we wonder if some other techniques could solve this drawback.

Since the performance of Grad-CAM depends on classifier models, we would ask:
\begin{itemize}
    \item What CNN architectures/types are good for segmentation by Grad-CAM? And why?
    \item How do the bottom layers (fully-connected layers, layers after the last convolutional layer) in CNN models affect the CAMs?
\end{itemize}

And for weakly-supervised image segmentation~\cite{kolesnikov_seed_2016,selvaraju2017grad}, CAMs can be referred and combined with other segmentation models, such as UNet CNNs~\cite{nguyen_novel_2019, rajapaksa_localized_2021}. More works could be done to fuse CAMs and segmentation models/methods to improve their performance.

\section{Conclusions}
In this study, we found the best segmentation from the Grad-CAM algorithm is 0.435 (averaged mean-Dice) by the Xception model. The Grad-CAM algorithm can recognize some of the important (tumor) areas to classification but its performance depends on classifier models. It indicates that the use of only Grad-CAM to train two-class CNN classifiers may not be an optimal approach for segmentation; instead, to combine Grad-CAM with other segmentation methods could be a promising direction. In addition, we have verified that the predictions of CNN classifiers mainly depend on tumor areas, and dark regions in Grad-CAM’s heatmaps also contribute to classification. We also proposed an evaluation method for the heatmaps; that is, to re-train and examine the performance of a CNN classifier that uses images filtered by heatmaps.
\hfill \break

\bibliography{ref}


\end{document}